\documentclass[preprint2]{aastex}

\newcommand{\lae}{\mathrel{<\kern-1.0em\lower0.9ex\hbox{$\sim$}}}
\newcommand{\gae}{\mathrel{>\kern-1.0em\lower0.9ex\hbox{$\sim$}}}
 
\shorttitle{IGR J16283--4838}
\shortauthors{Beckmann et al.}

\begin{document}

\title{{\it Swift}, {\it INTEGRAL}, {\it RXTE}, and {\it Spitzer} reveal IGR J16283--4838}


\author{V. Beckmann\altaffilmark{1}}
\affil{NASA Goddard Space Flight Center, Exploration of the Universe Division, Greenbelt, MD 20771, USA}
\email{beckmann@milkyway.gsfc.nasa.gov}
\and
\author{J. A. Kennea\altaffilmark{2}, C. Markwardt\altaffilmark{3,4}, A. Paizis\altaffilmark{5,6}, S. Soldi\altaffilmark{5}, J. Rodriguez\altaffilmark{7,5}, S. D. Barthelmy\altaffilmark{3}, D. N. Burrows\altaffilmark{2}, M. Chester\altaffilmark{2}, N. Gehrels\altaffilmark{3}, N. Mowlavi\altaffilmark{5,8}, J. Nousek\altaffilmark{2}}

\altaffiltext{1}{also with the Joint Center for Astrophysics, Department of Physics, University of Maryland, Baltimore County, MD 21250, USA}
\altaffiltext{2}{Department of Astronomy and Astrophysics, Pennsylvania State 
University, 525 Davey Laboratory, University Park, PA 16802, USA}
\altaffiltext{3}{NASA Goddard Space Flight Center, Exploration of the Universe Division, Greenbelt, MD 20771, USA}
\altaffiltext{4}{Department of Astronomy, University of Maryland, College Park, MD 20742, USA}
\altaffiltext{5}{INTEGRAL Science Data Centre, Chemin d' \'Ecogia 16, 1290
 Versoix, Switzerland}
\altaffiltext{6}{CNR-IASF, Sezione di Milano, via Bassani 15, 20133 Milano, Italy}
\altaffiltext{7}{Centre d'Etudes de Saclay, DSM/DAPNIA/SAp (CNRS UMR AIM/7158), 91191 Gif-sur-Yvette Cedex, France}
\altaffiltext{8}{Observatoire de Gen\`eve, 51 chemin des Maillettes, 1290 Sauverny, Switzerland}

\begin{abstract}
We present the first combined study of the recently discovered source \mbox{IGR~J16283--4838} with {\it Swift}, {\it INTEGRAL}, and {\it RXTE}. The source, discovered by {\it INTEGRAL} on April 7, 2005, shows a highly absorbed (variable $N_{\rm H} = 0.4 - 1.7 \times 10^{23} \, \rm cm^{-2}$) and flat ($\Gamma \sim 1$) spectrum in the {\it Swift}/XRT and {\it RXTE}/PCA data. No optical counterpart is detectable ($V > 20 \rm \, mag$), but a possible infrared counterpart within the {\it Swift}/XRT error radius is detected in the {\it 2MASS} and {\it Spitzer}/GLIMPSE survey.
The observations suggest that \mbox{IGR~J16283--4838} is a high mass X-ray binary containing a neutron star embedded in Compton thick material. This makes \mbox{IGR~J16283--4838} a member of the class of highly absorbed HMXBs, discovered by {\it INTEGRAL.}
\end{abstract}




\section{Introduction}

Star formation in our Galaxy takes place mainly in the dense regions of the spiral arms. These regions host massive molecular clouds and also the majority of the single and binary neutron stars ($\sim 10^9$) and black holes ($\sim 10^8$) in the Milky Way. The dense molecular clouds lead to strong star formation activity, which also results in the formation of binary systems, and subsequently to X-ray binary systems. These objects show X-ray flares and outbursts because of accretion processes onto the compact object. At the same time, the gas and dust of the spiral arms absorb most of the emission in the optical to soft X-ray regime below 10 keV. In addition, dense absorbing atmospheres around the object make the detection of these sources even more difficult. The hard X-ray and soft gamma-ray mission {\it INTEGRAL} \cite{INTEGRAL} operates at energies above 20 keV. With the large field of view of the main instruments, the imager IBIS (Ubertini et al.~2003; $19^\circ \times 19^\circ$, partially coded FOV) and the spectrograph SPI (Vedrenne et al.~2003; $35^\circ \times 35^\circ$, PCFOV), and its observing program focussed on the Galactic plane and center, {\it INTEGRAL} is a powerful tool to discover highly absorbed sources ($N_{\rm H} > 10^{23} \rm \, cm^{-2}$) in the Galactic plane. So far a handful of those enigmatic objects has been found since the launch of {\it INTEGRAL} in October 2002\footnote{for a list of all sources found by {\it INTEGRAL} see http://isdc.unige.ch/$\sim$rodrigue/html/igrsources.html}. Six of those sources have been published so far: IGR~J16318--4848 \cite{IGRJ16318} with an absorption of $N_{\rm H} \simeq 19 \times 10^{23} \rm \, cm^{-2}$ \cite{Matt}, IGR~J19140+0951 ($N_{\rm H} = 0.3 - 1.0 \times 10^{23} \rm \, cm^{-2}$; Rodriguez et al. 2005), IGR~J16320--4751 ($N_{\rm H} \simeq 2 \times 10^{23} \rm \, cm^{-2}$; Rodriguez et al.~2003), IGR~J16393--4643 ($N_{\rm H} \simeq 10^{23} \rm \, cm^{-2}$, Combi et al.~2004), IGR~J16358--4726 ($N_{\rm H} \simeq 4 \times 10^{23} \rm \, cm^{-2}$, Patel et al.~2004), and IGR~J16479--4514 ($N_{\rm H} > 5 \times 10^{23} \rm \, cm^{-2}$, Walter et al.~2004). While the nature of the latter source is still unknown, the other sources appear to be HMXBs, probably hosting a neutron star as the compact object. Most, if not all,  of these sources show variable absorption.
In this paper we report the discovery and analysis of another highly absorbed source, IGR~J16283--4838 \cite{INTEGRALatel1}. This work makes the first use of the combined data of {\it INTEGRAL}, {\it Swift}, {\it RXTE}, and {\it Spitzer}.
 
\section{Observations of IGR J16283--4838}

All observations discussed in this Section are summarized in Table~\ref{observations}.

\subsection{Discovery by {\it INTEGRAL}}

IGR~J16283--4838 was discovered \cite{INTEGRALatel1} during the observation of the Norma arm region by the imager IBIS/ISGRI \cite{ISGRI} on-board {\it INTEGRAL}. The observation 
lasted from April 7, 2005, 13:57 U.T. until April 9, 4:44 U.T. with an effective ISGRI exposure time of 126 ksec. The source position is $R.A. = 16^{\rm h}28.3'$, $DEC = -48^{\circ}38'$ (J2000.0) with 3 arcmin uncertainty. The source showed a flux of $f_X = (4.8 \pm 0.8) \times 10^{-11} \rm \, erg \, cm^{-2} \, s^{-1}$ in the 20 - 60 keV band. No emission was detectable above 60 keV. From the analysis of another ISGRI observation with similar exposure time we estimate the $3\sigma$ upper limit in the $60 - 200 \rm \, keV$ band $f_X < 1.2 \times 10^{-10} \rm \, erg \, cm^{-2} \, s^{-1}$.
The analysis of the data prior to the discovery 
lasting from April 4, 01:55 U.T. until April 6, 11:24 U.T., with an exposure time of 192 ksec resulted in a $3 \sigma$ upper limit of $f_{20 - 60 \rm \, keV} = 1.7 \times 10^{-11} \rm \, erg \, cm^{-2} \, s^{-1}$. 
The source showed significant brightening during an {\it INTEGRAL} observation
starting on April 10, 1:26 U.T. Even though IGR~J16283--4838 was in the partially coded field of view of IBIS, the analysis gave a $11.6 \sigma$ detection within 96 ksec with a flux of $f_{20 - 60 \rm \, keV} = (11.3 \pm 1.0) \times 10^{-11} \rm \, erg \, cm^{-2} \, s^{-1}$ \cite{INTEGRALatel2}. 
The low flux level of the source did not allow the extraction of a spectrum from the ISGRI data and no simultaneous soft-X and optical observations are available as IGR~J16283--4838 was always outside the field of view of {\it INTEGRAL}'s X-ray monitor JEM-X and of the optical monitor OMC.
No further {\it INTEGRAL} observations of the source were obtained. 

\subsection{X-ray follow-up observations}

After the discovery of IGR~J16283--4838 a {\it Swift} follow-up observation was requested in order to obtain an X-ray spectrum and an optical measurement. 
The {\it Swift} mission \cite{SWIFT} is a multiwavelength observatory for gamma-ray-burst astronomy. The payload combines a gamma-ray instrument (Burst Alert Telescope, 15 - 150 keV; Barthelmy et al.~2005), an X-ray telescope (XRT; Burrows et al.~2005), and a UV-optical telescope (UVOT; Roming et al.~2005). The XRT is a focussing X-ray telescope with a $110 \rm \, cm^2$ effective area, 23 arcmin FOV, 18 arcsec resolution, and 0.2-10 keV energy range. The UVOT design is based on the Optical Monitor (OM) on-board ESA's XMM-Newton mission, with a field of view of $17 \times 17 \rm \, arcmin$ and an angular resolution of 2 arcsec.

Two {\it Swift} observations took place 3 and 5 days after the last {\it INTEGRAL} observation. The first one started on April 13, 14:02 U.T. with an exposure time of 2.5 ksec, which resulted in an effective {\it Swift}/XRT exposure of 550 sec.  
A preliminary analysis of the XRT data refined the position of IGR~J16283--4838 to $R.A. = 16^{\rm h}28' 10.7''$, $DEC = -48^{\circ} 38' 55''$ (J2000.0) with an estimated uncertainty of 5 arcsec radius \cite{SWIFTatel}. 
A second observation was performed on April 15, 00:16 U.T. with 2600 sec effective XRT exposure time.

\begin{figure}
\plotone{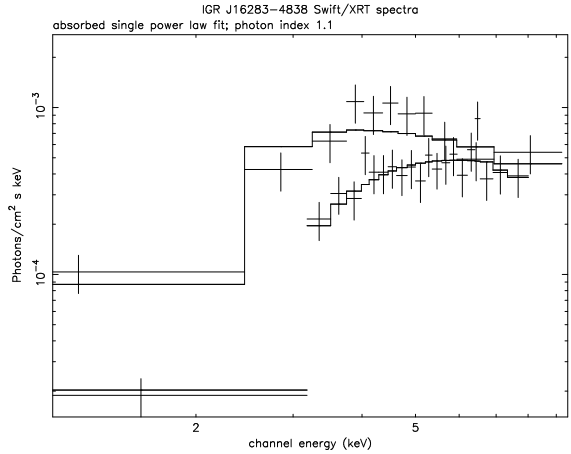}
\caption[]{{\it Swift}/XRT photon spectra of April 13 (upper spectrum) and of April 15 (lower spectrum). The applied fit is an absorbed power-law with $N_{\rm H} = 0.6 / 1.7 \times 10^{23} \rm \, cm^{-2}$ (April 13/15) and $\Gamma = 1.1$.}

\label{fig:XRTspectrum}
\end{figure}

For our analysis of the {\it Swift} data we used the calibration files which have been released on April 5, 2005 and the software provided by the Swift Science Center. These tools are included in the release of HEAsoft 6.0 as of April 12, 2005. 
Applying a centroid algorithm to the data of April 15 gives a refined position for the source of $R.A. = 16^{\rm h}28' 10.56''$, $DEC = -48^{\circ} 38' 56.4''$ with an uncertainty of 6 arcsec radius, consistent with both the preliminary analysis and the {\it INTEGRAL} measurement.
The spectrum extracted from the XRT data of April 15 is shown in Figure~\ref{fig:XRTspectrum}. The spectral fitting was done using version 11.3.2 of XSPEC \cite{XSPEC}. 
Both XRT spectra are well represented by an absorbed power law with the same photon index ($\Gamma = 1.12 \pm 0.35$), but different absorption column density. The observation of April 13 shows a less absorbed ($N_{\rm H} = 0.6 {+0.4 \atop -0.2} \times 10^{23} \rm \, cm^{-2}$) spectrum with a lower flux ($f_{\rm 2-10 \, keV} = (3.9 \pm 0.3) \times 10^{-11} \rm \, erg \, cm^{-2} \, s^{-1}$) than the April 15 one.  
The latter data show $N_{\rm H} = 1.7 {+0.5 \atop -0.4} \times 10^{23} \rm \, cm^{-2}$ and a flux in the 2 -- 10 keV band of $f_{\rm 2-10 \, keV} = (2.7 \pm 0.3) \times 10^{-11} \rm \, erg \, cm^{-2} \, s^{-1}$. 
The data are equally well fit by an absorbed black body with $N_{\rm H} = 0.3/1.4 \times 10^{23} \rm \, cm^{-2}$ (April 13/15) with a temperature of $kT = 2.0 \pm 0.3 \rm \, keV$.
Adding a Gaussian line to the fit does not improve the results significantly. The $3 \sigma$ upper limit for the Fe K$\alpha$ line at 6.4 keV is $3 \times 10^{-4} \rm \, ph \, cm^{-2} \, s^{-1}$ with an equivalent width of $EW < 600 \rm \, eV$. 
Because of the short exposure time the source was not detected by the BAT instrument.

IGR J16283--4838 was then also observed twice by {\it RXTE} using the Proportional Counter Array (PCA). The first observation starting on April 14, at 0:46 U.T. lasted 3.6 ksec, the second on April 15, at 16:07 U.T. lasted 2.9 ksec \cite{RXTEatel}. During both observations the PCA pointing was offset by 45 arcmin to avoid the nearby bright low mass X-ray binary 4U~1624--490.
The {\it RXTE}/PCA has a large field of view ($2^\circ$ FWZM). For targets
near the Galactic plane, a significant amount of Galactic diffuse
emission enters the PCA aperture, which is considered background.
This background was modeled by taking a nearby observation of the
Galactic plane (observation 91409-01-02-00, $l=341.4^\circ$,
$b=0.6^\circ$). This observation is at a similar latitude as
IGR~J16283--4838, so the diffuse emission should have nearly the same
spectrum. The background observation was modeled as thermal
bremsstrahlung with a temperature of $kT = 7.4$ keV, plus line
emission at $\sim 6.5$ keV with an equivalent width of 600 eV. The
shape of the background template was fixed and added to the spectral
model of the two PCA observations of IGR~J16283-4838; only the total normalization of the template was allowed to vary. The fluxes are collimator corrected after background subtraction. The best fit models for the source are shown in Table~\ref{observations}.


No pulsations are detectable in the PCA data.

\subsection{Infrared and optical data}

Within the 6 arcsec error radius around the refined position determined from the {\it Swift}/XRT data the infrared source 2MASS J16281083--4838560 is located at 2.7 arcsec distance (Rodriguez \& Paizis 2005). This source has K, J, and H band magnitudes of $K = (13.95 \pm 0.06) \rm \, mag$, $H > 15.8 \rm \, mag$, and $J > 16.8 \rm \, mag$ ($95 \%$ lower limits).

The Galactic Legacy Infrared Midplane Survey Extraordinaire (GLIMPSE\footnote{publicly available at
http://irsa.ipac.caltech.edu/data/SPITZER/GLIMPSE/}; Benjamin et al.~2003) data show the source SSTGLMC G335.3268+00.1016 at 2.9 arcsec distance to the XRT position, consistent with the 2MASS detection. GLIMPSE is a 4-band near- to mid-infrared survey by {\it Spitzer} \cite{Spitzer} of the inner two-thirds of the Galactic disk with a spatial resolution of $\sim 2''$. The Infrared Array Camera \cite{IRAC} imaged 220 square degrees at wavelengths centered on 3.6, 4.5, 5.8, and 8.0 $\mu m$ in the Galactic longitude range $10^\circ$ to $65^\circ$ on both sides of the Galactic center and in Galactic latitude $\pm 1^\circ$. The {\it Spitzer}/GLIMPSE data show a clear detection 
in all four energy bands (Tab.~\ref{observations}).
Another observation in the K-band was performed with the 6.5m Magellan-Baade telescope on April 21, 2005. This observation indicates that the source seen in the 2MASS is a blend of point sources, with the brightest showing $K = 14.1\rm \, mag$ \cite{newIR}. Therefore the identification with the {\it Spitzer} source is tentative. In case the infrared source is not the counterpart to the hard X-ray source, the data presented here would be upper limits for the near and mid-infrared emission.  

\begin{figure}
\plotone{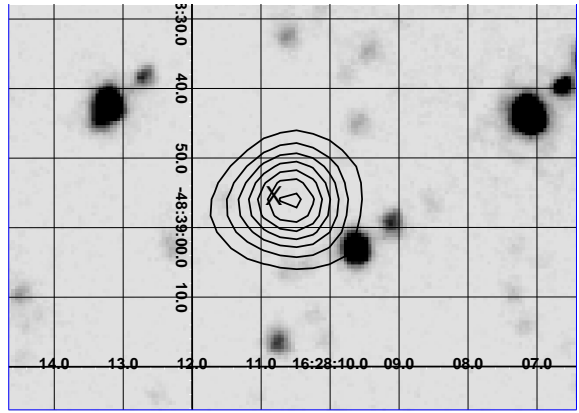}
\caption[]{{\it Swift}/XRT contour plot ontop of the UVOT V-band map of IGR~J16283--4838 based on a 2.6 ksec observation on April 15, 2005. The cross indicates the position of the mid-infrared source SSTGLMC G335.3268+00.1016 seen by {\it Spitzer} and in the {\it 2MASS}.}

\label{fig:XRTimage}
\end{figure}

Within the error radius of IGR~J16283--4838 no optical counterpart is detectable on the POSS-II plates of the Digitised Sky Survey. 
During the observations by {\it Swift} on April 13 and on April 15 the UVOT took an image in the V-band. No source is detected within the error radius down to a magnitude of $V > 20 \rm \, mag$.      
The image extracted from the {\it Swift}/UVOT data on April 15 is shown in Figure~\ref{fig:XRTimage}. The contours indicate the XRT count map and the cross
gives the position of the mid-infrared counterpart.

\section{Spectral Energy Distribution}

\begin{figure*}
\plotone{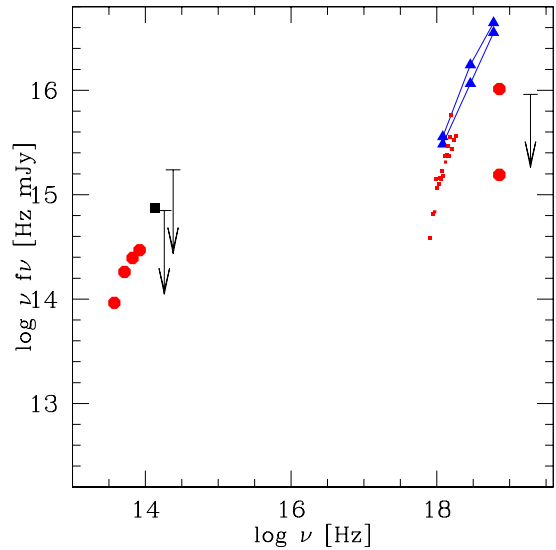}
\caption[]{Spectral energy distribution of IGR~J16283--4838. From the left, {\it Spitzer}/GLIMPSE (marked by octagons), Magellan-Baade (square), {\it 2MASS} (two upper limits),
{\it Swift}/XRT (small dots), 
and marked by triangles the {\it RXTE}/PCA data (upper line: April 13, lower line: April 15). The two octagons on the right 
represent the {\it INTEGRAL}/ISGRI measurement on April 8 (low flux) and April 10 (higher flux). The arrow on the right shows the ISGRI upper limit at 80 keV. Unabsorbed fluxes are displayed considering the Galactic but not the intrinsic absorption.}

\label{fig:SED}
\end{figure*}

The spectral energy distribution (SED) of IGR~J16283--4838 is shown in Figure~\ref{fig:SED}. In the chosen diagram a single power law with photon index $\Gamma = 2$ would appear as an even, horizontal line. 
No error bars have been included for the {\it Swift}/XRT data and only the XRT data of April 15 are shown for better visibility.
From the comparison of the XRT data points with the measurements by {\it RXTE}/PCA it is apparent that both were taken during a similar high state of the source, while the two {\it INTEGRAL}/ISGRI measurements describe a lower flux state. Unfortunately the $60 - 200 \rm \, keV$ upper limit does not constrain the SED significantly. 

Note that we display in the SED the absorbed X-ray fluxes as they are measured at the observer, as most of the absorption appears to be intrinsic to the source.  
The situation is different in the optical where the flux is absorbed already significantly by material in the line of sight. The hydrogen column density in the direction of the source is $N_{\rm H} = 2.2 \times 10^{22} \rm \, cm^{-2}$. This leads to an extinction of $A_V = N_{\rm H} / 1.79 \times 10^{21} \rm \, cm^{-2} = 12.3 \rm \, mag$ \cite{extinction}. Therefore the unabsorbed optical limit is $V > 7.7 \rm \, mag$ and outside the displayed range of Figure~\ref{fig:SED}. The absorption has a lower effect on the near-infrared fluxes. With $A_K = 0.112 \, A_V$, $A_H = 0.176 \, A_V$, and $A_J = 0.276 \, A_V$ (Schlegel, Finkbeiner, \& Davis 1998) the unabsorbed flux values are $K = 12.7  \rm \, mag$, $H >  13.6 \rm \, mag$, and $J > 13.4 \rm \, mag$. The extinction in the mid-infrared range is negligible. 

\section{Discussion}

The new hard X-ray source IGR~J16283--4838 exhibits several characteristic features.
IGR J16283--4838 is located at Galactic longitude $l = 335.3^\circ$ and latitude $+6.1$ arcmin in the Norma arm region. 
It shows a strong flare within a time scale of days. 
The absorption is of the order of $N_{\rm H} = 0.4 - 1.7 \times 10^{23} \rm \, cm^{-2}$ with variations by a factor of 4 within one day and a flat X-ray spectrum ($\Gamma \simeq 1$) during all observations.
The bimodal spectral energy distribution has one peak probably in the near-infrared and the other in the hard X-rays.
The equivalent width of the iron K$\alpha$ line is $EW < 600 \rm \, eV$ ($f_{\rm K \alpha} < 3 \times 10^{-4} \rm \, ph \, cm^{-2} \, s^{-1}$)
and no optical counterpart is detectable ($V > 20 \, \rm mag$), probably because of absorption in the line of sight.

Combining this information enables us to put constraints on the nature of the source. 
The position within the Galactic plane at only $+6.1$ arcmin makes a Galactic origin of the source likely, even though some AGN have been seen through the plane by {\it INTEGRAL}, like the Seyfert 1 galaxy GRS 1734-292 \citep{GRS1734a,GRS1734b}. Strong variability as observed in IGR~J16283--4838 has been seen in the X-ray spectra of Seyfert galaxies (e.g. Dewangan et al. 2002), but the X-ray spectrum is too flat ($\Gamma \simeq 1$) for a Seyfert galaxy. 
This would still leave the possibility of a blazar as counterpart.
But the absorption by the Galaxy in the direction of the source ($N_{\rm H} = 2.2 \times 10^{22} \rm \, cm^{-2}$) is not high enough to explain the intrinsic absorption of $1.7 \times 10^{23} \rm \, cm^{-2}$, and thus intrinsic strong absorption in the blazar would be required to explain the {\it Swift}/XRT spectrum, but this has not been seen so far in blazar spectra. 

If we consider IGR~J16283--4838 to be a Galactic source, mainly two types of bright and variable hard X-ray emitters are likely to be a counterpart: Low Mass and High Mass X-ray Binaries, LMXBs and HMXBs, respectively. 
The hard X-ray spectrum with strong absorption indicates the presence of a HMXB in which no pulsation have been detected so far \cite{RXTEatel}. Also the bright infrared emission, if connected to the X-ray source, would indicate a massive star as the companion of the compact object. 

For a HMXB it is likely that IGR~J16283--4838 is located close to a star forming region in a Galactic spiral arm. Along the line of sight towards the source several arms are located \cite{arms}: the Sagittarius-Carina arm (0.7 kpc), the Scutum-Crux arm (3.2 kpc), the Norma-Cygnus arm (4.8 kpc), a star-forming region (7 kpc), and the Perseus arm (10.8 kpc). 
The luminosity of the object during the flare can be estimated by taking the brightest stage during the {\it RXTE} observation and assuming a distance to the object between 1 and 10 kpc. The unabsorbed flux is in this case only 20\% larger than the absorbed one, because the significant part of the luminosity is emitted in the hard X-rays. The bolometric luminosity is then in the range $\log L_{\rm burst} = 34.0 - 36.5$ (where $L$ is in units of $\rm \, erg \, s^{-1}$). The quiescent luminosity of the system is at least a factor of $\sim 20$ lower with $\log L_{\rm q} < 33 - 35.2$.
This range of values is consistent with measurements from known Be/X-ray binaries with a neutron star as the compact object \cite{BeX}.
 In any case the luminosity is far below the Eddington luminosity of a neutron star of $1.4 M_\odot$ ($L = 1.8 \times 10^{38} \rm \, erg \, s^{-1}$).  

The properties of IGR~J16283--4838 are similar to those of a number of highly absorbed sources ($N_{\rm H} = 1 - 20 \times 10^{23} \rm \, cm^{-2}$) found in the Galactic plane, especially in the Norma arm region \cite{IGRs}. The HMXB IGR~J19140+0951 shows also strong variable absorption \cite{IGRJ19140}, indicating intrinsic absorption in the source. The observed properties of IGR~J16283--4838 are consistent with those of IGR~J19140+0951 in the bright state, where the iron line flux decreased to $4 \times 10^{-4} \rm \, ph \, cm^{-2} \, s^{-1}$, which is at the upper limit for the {\it Swift}/XRT measurement in our case. 
The (non-)variability of the absorption in IGR~J16318--4848 is still under discussion, as Walter et al. (2003) claim constant absorption, whereas Revnivtsev (2003) discovered variable absorption which could be connected with the orbital phase of the binary system.
Only one of the newly detected highly absorbed sources has been claimed so far not to be a HMXB. Patel et al.~(2004) observed IGR~J16358--4726 with {\it Chandra}. From the X-ray data they favour the source to be a millisecond pulsar LMXB even though the HMXB interpretation cannot be ruled out completely, though it would require some unknown kind of spin-down torque to prevent the neutron star from spinning up in this particular case.

X-ray binaries with strong intrinsic absorption have been known already before {\it INTEGRAL}, for example in 4U 1700-377, GX 301-2, Vela X-1, and CI Cam. Except for the latter one, where the nature of the source is unclear to date, these sources are also HMXBs, likely to host a neutron star as the compact object. Vela X-1 shows variable absorption from a negligible value up to $7 \times 10^{23} \rm \, cm^{-2}$ \cite{VelaX-1}. GX 301-2 shows strong absorption variation (up to $12 \times 10^{23} \rm \, cm^{-2}$; White \& Swank~1984), and so does CI Cam ($(0.02 - 5) \times 10^{23} \rm \, cm^{-2}$; Boirin et al.~2002). In 4U 1700-377 the absorption is linked to the state of the HMBX system and varies by a factor of 2 between 0.9 and $2.0 \times 10^{23} \rm \, cm^{-2}$ \cite{4U1700}. It appears that variable absorption is a common feature in highly absorbed HMXBs.  
This could mean that the absorbing material is linked to the existence of a high mass donor in the binary system. In this case a strong and dense stellar wind ($10^{-7}$ to $10^{-5} \, M_\odot \rm \, yr^{-1}$) from the early-type stellar companion will probably cause the absorption in the system. 
The fact that all the absorbed sources so far have shown to be HMXBs \citep{kuulkers,IGRs} containing neutron stars does not rule out significant contribution of HMXBs with a black hole as the compact object. But these systems are expected to be less numerous than the neutron star HMXBs by a factor of 10 to 100, making the detection of a black hole binary within a sample of only about 10 detected highly absorbed HMXBs unlikely. 
%
These absorbed binary systems might provide a significant contribution to the Galactic hard X-ray background at energies above 10 keV \citep{XRB2,XRB}.

\section{Conclusions}

The newly discovered hard X-ray source \linebreak \mbox{IGR~J16283--4838}, located in the Norma arm region, is likely to be a HMXB containing a neutron star as the compact object. It is located in the Galactic Plane in the direction of star forming regions in the spiral arms and shows a large flare, which makes an extragalactic origin unlikely. The spectrum is hard ($\Gamma \sim 1$) and strongly absorbed during the flare, which indicates a HMXB rather than a LMXB.  The luminosity is comparably low ($L < 10^{37} \rm \, erg \, s^{-1}$) which is typical for a neutron star HMXB. 
The strong and variable absorption ($N_{\rm H} = 0.4 - 1.7 \times 10^{23} \rm \, cm^{-2}$) indicates that IGR~J16283--4838 belongs to the class of highly absorbed HMXBs discovered by {\it INTEGRAL} along the Galactic plane.
Bright and absorbed sources like IGR~J16283--4838 could contribute significantly to the Galactic hard X-ray background in the 10--200 keV band.

It has to be pointed out that the discovery and classification of IGR~J16283--4838 would not be possible without combining the observations of the recent observatories in space, like {\it INTEGRAL}, {\it Swift}, {\it RXTE}, and {\it Spitzer}. Combined efforts from these missions should lead to deeper insights into the nature of the hard X-ray source population in our Galaxy in the near future. 
 
\begin{acknowledgements}
We like to thank John Greaves for pointing out the GLIMPSE data. 
This work is based in part on observations made with the Spitzer Space Telescope, which is operated by the Jet Propulsion Laboratory, California Institute of Technology under NASA contract 1407. This research has made use 
of the SIMBAD Astronomical Database which is operated by the Centre de Donn\'ees astronomiques de Strasbourg. This work was supported in part by NASA 
contract NAS5-00136.
\end{acknowledgements}

%
%
\begin{table*}
\caption[]{Summary of observations}
\begin{tabular}{lccccc}
\tableline\tableline
Instrument & Date & Energy range & Flux$^a$ & $N_H$  & $\Gamma$ \\
     & &  &  & $10^{22} \, \rm cm^{-2}$ & \\    
\hline                        
{\it Spitzer}        & 2004  & 3.6 $\rm \mu m$ & $3.5 \pm 0.2$ mJy & - & - \\
 & 2004      & 4.5 $\rm \mu m$ & $3.7 \pm 0.3$ mJy  & - & - \\
 & 2004      & 5.8 $\rm \mu m$ & $3.5 \pm 0.4$ mJy  & - & - \\
 & 2004      & 8.0 $\rm \mu m$ & $2.5 \pm 0.2$ mJy  & - & - \\
{\it 2MASS}          & June 18, 1999 & K band      & $13.95 \pm 0.06$ mag & - & - \\
                 & June 18, 1999 & J band      & $> 16.8$ mag & - & - \\
                 & June 18, 1999 & H band      & $> 15.8$ mag & - & - \\
Magellan-Baade & April 21 & K band & 14.1 mag & - & - \\
{\it Swift}/UVOT & April 13/15  & V band      & $> 20$ mag & - & - \\ 
{\it Swift}/XRT  & April 13  & 2--10 keV    & $3.9 \pm 0.3$  & $6 \pm 2$  & $1.1 \pm 0.4$ \\
 & April 15  & 2--10 keV    & $2.7 \pm 0.3$  & $17 \pm 4$  & $1.1 \pm 0.4$ \\ 
{\it RXTE}/PCA       & April 14  & 2--10 keV    & $5.8 \pm 0.3$   & $13 \pm 6$
 & $0.9 \pm 0.1$\\ 
   &           & 10--20 keV   & $13.2 \pm 0.7$    &   &    \\ 
   &           & 20-40 keV    & $30.7 \pm 1.5$    &   &    \\ 
   & April 15  & 2--10 keV    & $4.9 \pm 0.7$   & $4 \pm 4$ & $0.8 \pm 0.3$ \\ 
   &           & 10--20 keV   & $8.8 \pm 1.3$     &   &    \\ 
   &           & 20--40 keV   & $20.6 \pm 3.1$    &   &    \\ 
{\it INTEGRAL}/ISGRI & April 4--6 & 20--60 keV   & $< 1.7 $         & - & -  \\   & April 7--9 & 20--60 keV   & $4.8 \pm 0.8$  & - & -  \\ 
   & April 10  & 20--60 keV   & $11.3 \pm 1.0$ & - & -  \\ 

\tableline
\end{tabular}
\label{observations}

$^{a}$ measured flux in $10^{-11} \rm \, erg \, cm^{-2} \, s^{-1}$ if not indicated differently
\end{table*}
%
%
%

\end{document}